\documentclass[%
 reprint,
superscriptaddress,
 amsmath,amssymb,
 aps,
]{revtex4-2}

\usepackage{graphicx}
\usepackage{dcolumn}
\usepackage{bm}
\usepackage{xr-hyper}
\usepackage{hyperref}
\hypersetup{
    colorlinks=true,
    linkcolor=blue,
    citecolor=blue,
    urlcolor=blue,
    }
\usepackage[mathlines]{lineno}
\usepackage{mathtools}
\usepackage{physics}
\usepackage{placeins}
\usepackage{multirow}
\usepackage{url}
\usepackage{xcolor}
\usepackage{soul}

\makeatletter
\newcommand*{\addFileDependency}[1]{
\typeout{(#1)}
\@addtofilelist{#1}
\IfFileExists{#1}{}{\typeout{No file #1.}}
}\makeatother

\newcommand*{\myexternaldocument}[1]{%
\externaldocument{#1}%
\addFileDependency{#1.tex}%
\addFileDependency{#1.aux}%
}
\newcommand{\HMMtwo}[1][]{in the main text.}
\myexternaldocument{HMMtwo}

\DeclareUnicodeCharacter{2212}{-}

\PassOptionsToPackage{numbers, compress}{natbib}

\begin{document}

\preprint{APS/123-QED}

\title{Unraveling cavity-like modes of two-dimensional broad band hyperbolic metamaterial and their coupling to quantum emitters}

\author{Amitrajit Nag}
\email{amitrajitnag@iisc.ac.in}
 \affiliation{
 Indian Institute of Science, C.V. Raman Road, Bangalore, India, 560012}
 \author{Girish S. Agarwal}
 \affiliation{
 Texas A\&M University, College Station, TX, USA, 77843}
 \author{Jaydeep K. Basu}
 \email{basu@iisc.ac.in}
 \affiliation{
 Indian Institute of Science, C.V. Raman Road, Bangalore, India, 560012}
%

\begin{abstract}
Hyperbolic metamaterials (HMM) are artificially engineered materials that exhibit hyperbolic dispersion of light propagating through them. These have been extensively studied for tailoring light propagation. Most studies use an effective medium approach that is extremely useful, though it misses out on properties that can arise from the microscopic details of the HMM. In particular, the HMM can have cavity-like modes, and it is important to understand such modes and their relevance in light propagation and coupling of HMM to quantum emitters. In this work, we bring out the cavity-like modes of the silver nanowire-alumina two-dimensional HMM, which remain on top of the broad response of the HMM. These modes define the characteristic reflection spectra. The observed resonances and their widths are in good agreement with our simulations. These well-defined modes occur even though the metallic part of the HMM has Ohmic losses. 
Then, we present experimental results on the coupling of quantum emitters to the cavity-like modes of the HMM. We present results for both steady-state and time-resolved photoluminescence. Using these, we extract the corresponding Purcell factors for radiative rate enhancement. Theoretical analyses of the experimental data allow the determination of the cavity coupling parameters and mode volumes. These experimental results are confirmed by the FDTD calculations for the HMM mode volume. This work elucidates the pathway to precise engineering for future applications of HMM modes in strong light-matter interactions.     
\end{abstract}

\maketitle



\section{Introduction}
Plasmonic materials attract attention in light-matter interactions due to their potential to confine electromagnetic field modes within subwavelength volume \cite{subwavelength_review, subwavelength_2, subwavelength_3, subwavelength_4} and enhance the field response with the help of their resonant coupling of the external electromagnetic field to the plasmon of the metallic nanostructure, generating the plasmon polaritons \cite{SPP_review, SPP_PRL}. Plasmonic materials are used for a myriad of applications in spontaneous emission engineering and optical field enhancements \cite{Plasmonic_enhancement, Pasmonic_enhancement2}, sensing \cite{Plasmonic_sensing, Plasmonic_sensing2}, energy harvesting devices \cite{Plasmonic_energyHarvest, Plasmonic_energyHarvest2}, unconventional optical microscopy \cite{Plasmonic_Microscopy, Plasmonic_microscopy2, Plasmonic_microscopy3}, and so on.
\begin{figure}[b]
    \centering
    \includegraphics[width=0.5\textwidth]{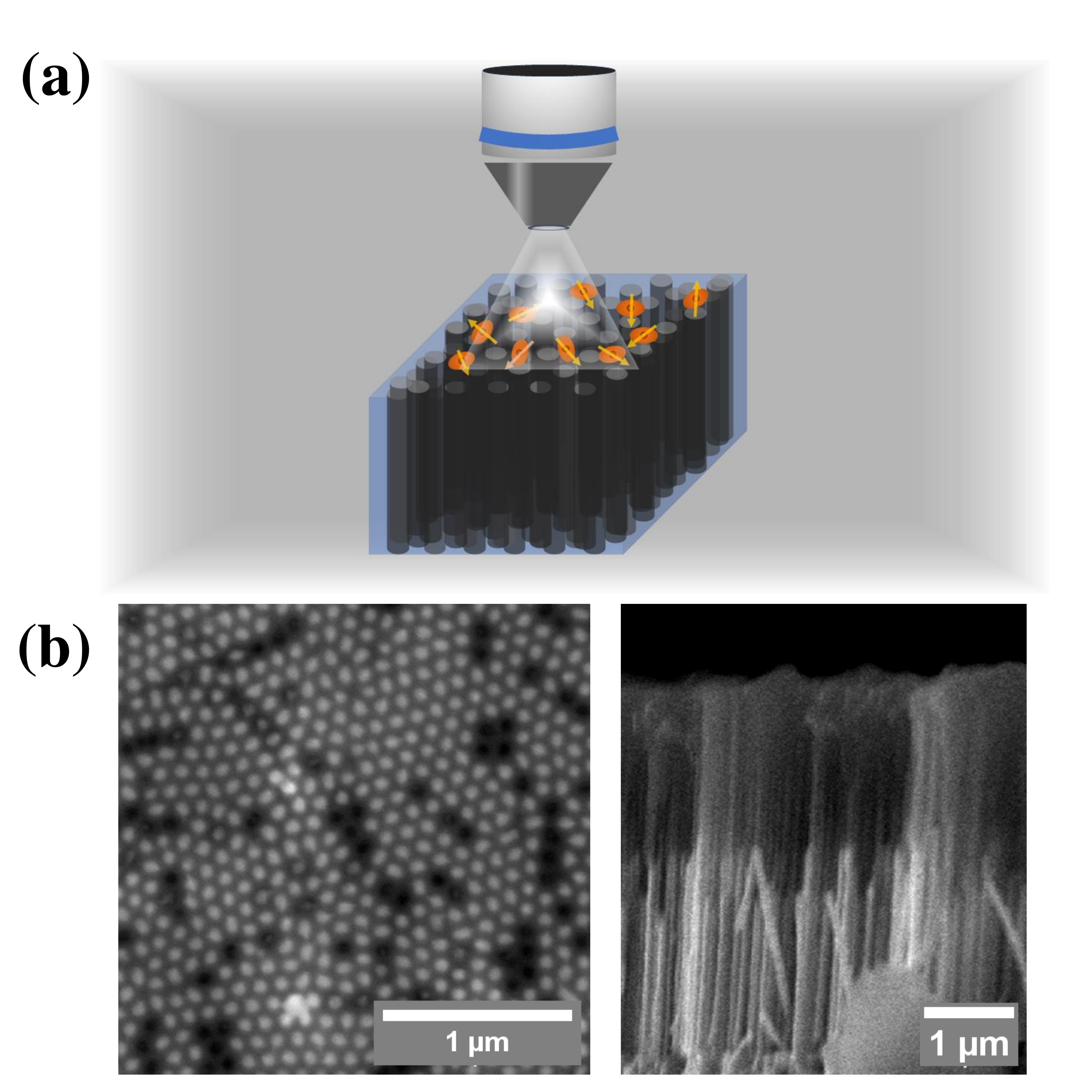}
    \caption{\textbf{(a)} Schematic of the modeled HMM system composed of AgNWs coupled to quantum emitters under excitation using an objective at the top for light-matter interaction studies. The same objective collects the scattering responses. \textbf{(b)} Scanning electron microscope (SEM) image of the AgNW-grown HMM sample from the top of the FIB etched sample (left) showing the shape of the triangular lattice, and the cross-section image shows the AgNWs grown inside the structure (right).
    }
    \label{fig:schematic}
\end{figure}
Artificially designed and engineered plasmonic metamaterials provide lucrative advantages upon subwavelength confinement volume for the electromagnetic field with easily tunable design parameters that are easier to fabricate than their all-dielectric counterparts. Hyperbolic metamaterials (HMM) made of plasmonic materials provide such unique properties because their isofrequency surface is a hyperboloid, ideally divergent since they are not bound and, hence, non-integrable. This ensures that the photonic local density of states (PDOS) for the available electromagnetic modes is also diverging, and propagating high wave vector modes within the metamaterial volume make the HMM system an impressive aspirant for extensive studies in light-matter interactions \cite{HMM_JubinZ, HMM_intro2, HMM_intro3, HMM_intro4_nonlocal}.
The large propagating wave vectors within the HMM volume immensely reduce the spatial dimensions for field confinements, which is a conquest for the system over other nanostructure systems of similar interest.
HMMs made of noble metal nanostructures are capable of showing all the qualities of being a plasmonic cavity system when used for super-resolution imaging and microscopy \cite{HMM_hyperlens, HMM_superresolution}, sensing \cite{HMM_biosensing, HMM_biosensing2, HMM_sensorAbsorber}, energy harvesting \cite{HMM_energyHarvest}, etc. Our group has reported spontaneous emission enhancement to strong coupling \cite{HMM_SRK, HMM_RKY2, HMM_Harsha} and photonic spin-momentum locking \cite{HMM_RKY} for light-matter interactions between the HMM and the ensemble of quantum emitters before. Thus, HMM has all the ingredients to emerge as a viable plasmonic cavity system. Theoretically, the quest of HMM cavity properties was addressed already, simulating silver nanorod HMM of square lattice arrangement \cite{PNAS_HMM_indefiniteCavity}.\\
This paper aims to explore the photonic indefinite medium cavity properties of the composite nanostructure of metal nanowire-dielectric HMM, which has not been experimentally studied before. We thrive in the quest using light-matter interaction experiments, followed by rigorous theoretical analyses and numerical simulations to support and explain the experimental findings.
We started our study with broadband excitation of the HMM system of silver nanowire (AgNW) arrays embedded in the alumina matrix (Al$_2$O$_3$) to learn its mode properties from reflectivity and absorption behaviors. Next, Rhodamine 6G dye (Rh6G) was dispersed over the HMM interface to study the light-matter interaction for HMM coupled to those dye quantum emitters; experiments were performed at room temperature (RT) as well as at a lower cryogenic temperature around 4K to reduce the scattering losses of HMM having a dissipative nature due to Ohmic losses \cite{AgNW_PhononScatt_SciRep, Ohmic_loss}. Steady-state and time-resolved Photoluminescence (PL \& TRPL) spectra were collected from the system, revealing modifications of spontaneous emission decay and Purcell enhancements. Subsequently, theoretical analyses of the density matrix master equations were performed to quantify the coupling strength with the help of decay parameters and cavity mode linewidth; also, we have used the finite difference time domain (FDTD) method to numerically simulate Maxwell's equations in the model periodic AgNW-HMM structure and record its responses; these results supported our experimental observations of mode linewidth and helped to quantify cavity mode volumes independently.

\section{Sample preparation}
The HMM was fabricated using chemical electrodeposition of AgNWs within the porous Al$_2$O$_3$ membrane, obtained from two-step anodization of an aluminum sheet, followed by a focused ion beam (FIB) etching to remove the top metal layer \cite{HMM_SRK, HMM_RKY, HMM_Harsha}; a dilute concentration of Rh6G dye was spread over the FIB-etched HMM interface and separated by a non-interacting polymer spacer layer. Figure \ref{fig:schematic}.(a) presents the schematic of the studied system; sparse randomly oriented model quantum emitters over the modeled block of AgNW-grown-HMM are excited with an objective from the topside, and the emission is collected through the same. Figure \ref{fig:schematic}.(b) shows the scanning electron microscopy (SEM) image of the FIB-etched HMM sample from the top and from the cross-section side. AgNWs are filled into the porous Alumina in a triangular lattice structure; the dark portions in the SEM images show the defect sites (left) and the unfilled top part (right). Further details and information on the measurement setup can be found in Appendix \ref{app:setup}.

\section{Results}
\subsection{Cavity-like modes of HMM}
Optical Topological Transition (OTT) of an HMM \cite{OTT1, OTT2} is estimated from the Maxwell-Garnett effective medium theory (EMT) applied for anisotropic materials \cite{EMT_MG_1, EMT_MG_2}; EMT only considers the metal filling fraction of the HMM to determine its behaviors. After the OTT point, HMM adopts its hyperbolic nature. The phenomenological density of states parameter $P$ can bridge the EMT parameters and the measured absorption properties of HMM to estimate its light-matter interaction behaviors \cite{HMM_Harsha, jacob2012broadband},
but EMT falls short of explaining any mode properties of the HMM system \cite{HMM_SRK, HMM_Harsha}, and so does $P$. Therefore, we need to go beyond EMT, where we estimate the mode properties of the HMM accurately. In this regard, we performed FDTD simulations on the periodic array structure of the model AgNW-HMM system under broadband plane wave excitation and recorded the scattering responses from the structure, which showed a frequency dispersion and resolved the HMM modes. In the simulation, AgNW diameter was set to 60 nm, periodicity was set to 120 nm, and periodic boundary conditions were applied along the plane of the HMM to emulate the array; these values correspond to the fabricated HMM system parameters obtained from SEM images and the metal filling fraction. FDTD simulation can explore the minute details of the system response due to the precision of the computation mesh, thus recording the cumulative effect of true nanostructure response rather than assuming it as a homogenized medium.\\ 
To understand the mode properties of the HMM, we need to conduct transmission measurements on it, but HMM is a plasmonic system with a heavy amount of losses present, so we cannot get a transmission signal through it. Therefore, we conduct the differential reflection spectroscopy with a broadband plane wave source exciting the bare HMM. This is capable of providing absorption properties for the HMM \cite{HMM_Harsha, Diff_Reflect_1}.
\begin{figure*}[!ht]
    \centering
    \includegraphics[width=0.8\textwidth]{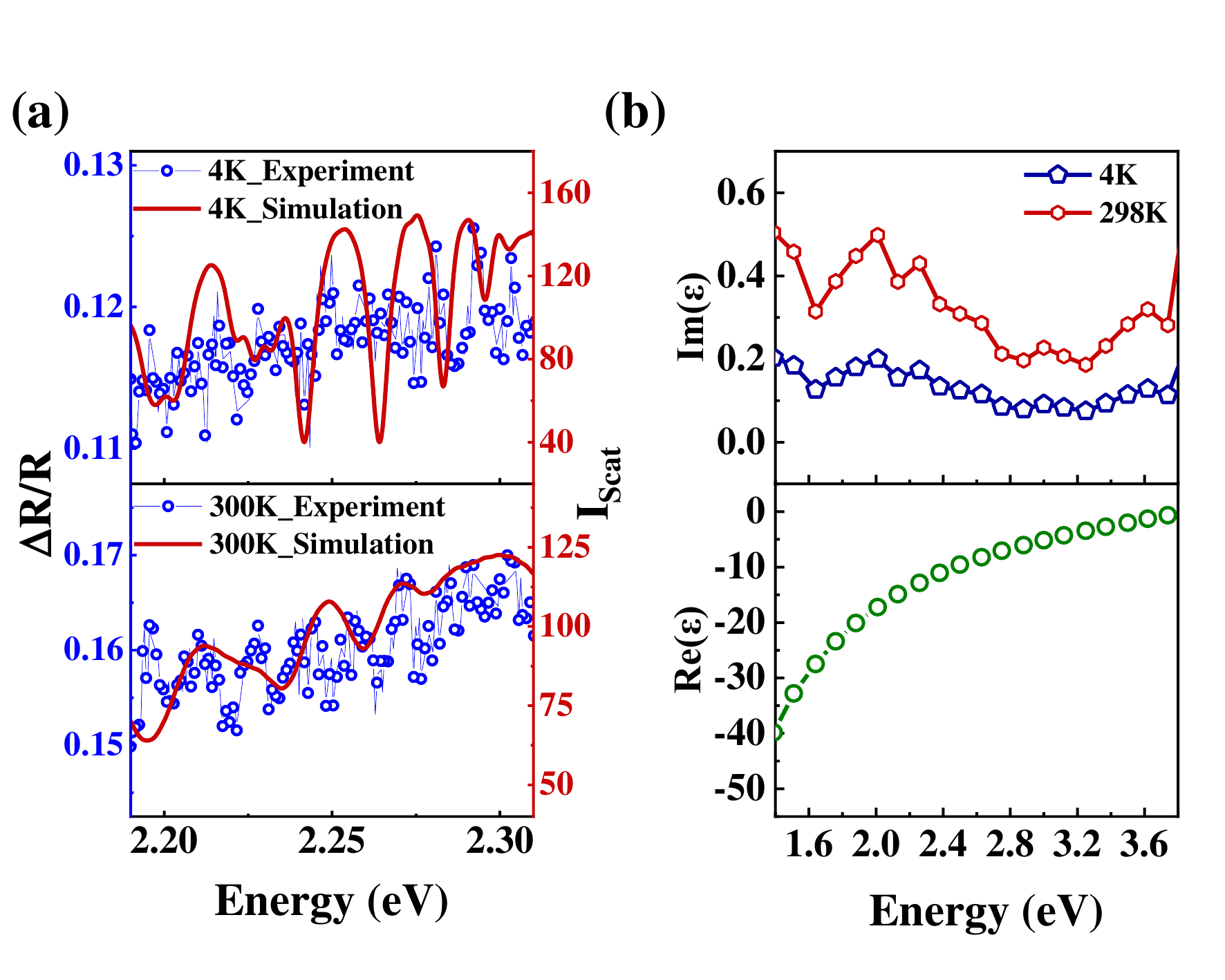}
    \caption{\textbf{(a)} The spectrum of the HMM modes obtained from the differential reflection spectroscopy and corresponding simulated scattering spectra of the HMM system under broadband plane wave excitation overlaid upon it shows mutual consistency. It is to be noted that modes obtained from differential reflection correspond to the dips in the spectra. The thin line (blue) connects the reflectivity data points.
    \textbf{(b)} Dielectric constants of silver are shown: the imaginary part at 4K and room temperature (RT) and the real part at RT. Only the imaginary part of the dielectric constant gets affected by the temperature; $Re(\epsilon)$ is temperature independent. $Im(\epsilon)$ data at room temperature is multiplied by the effective damping constant to yield the respective data at corresponding lower temperatures.}
    \label{fig:FDTD_WL}
\end{figure*}

\begin{figure*}[!ht]
    \centering
    \includegraphics[width=0.8\textwidth]{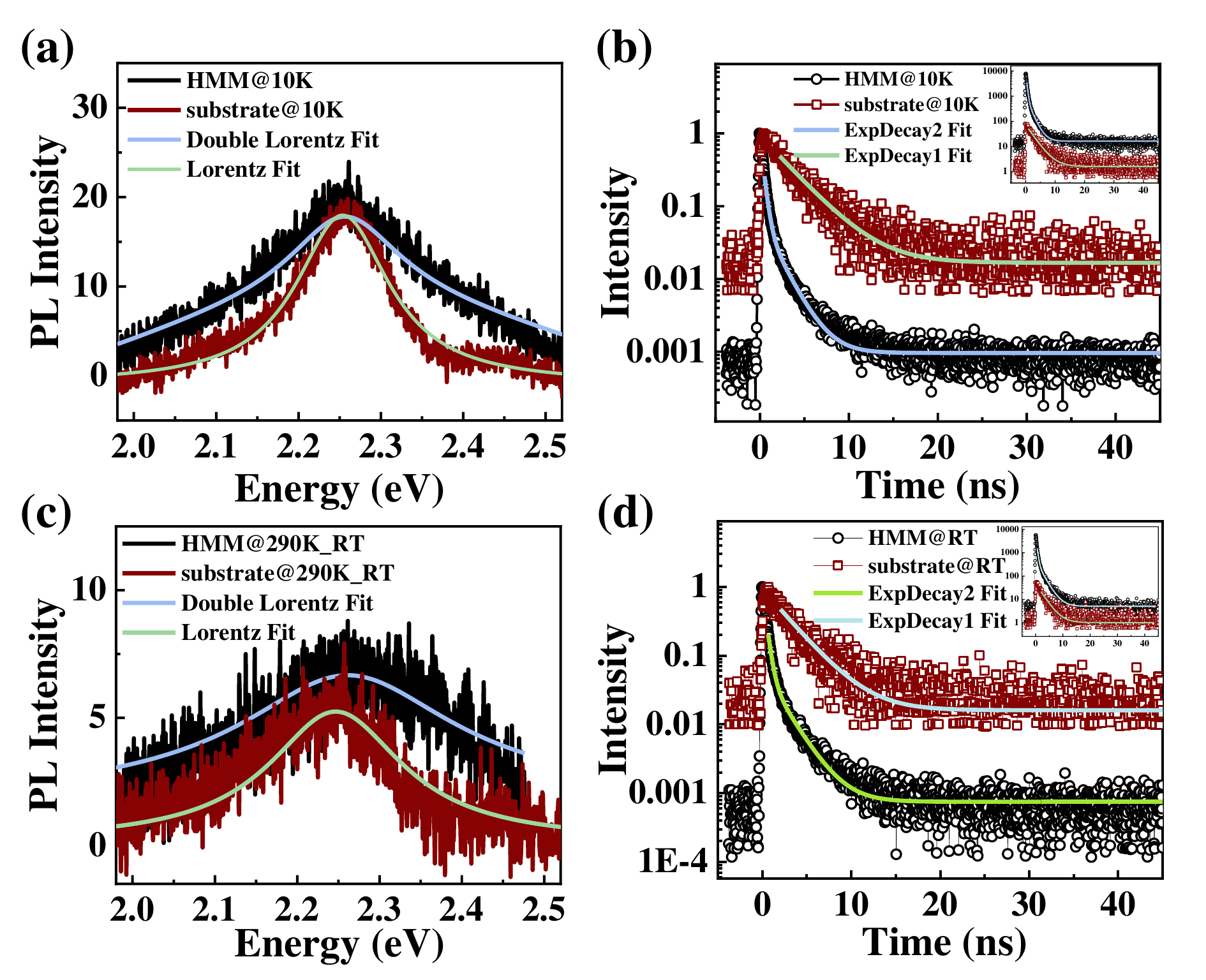}
    \caption{\textbf{(a)} PL \& \textbf{(b)} TRPL spectra of Rh6G coupled to the HMM system (black) and corresponding spectra of the dye over the substrate (red). (a) presents PL enhancement, and (b) shows faster spontaneous emission decay from the HMM than the substrate. The inset shows unnormalized data depicting enhancement for emission from HMM. \textbf{(c)} \& \textbf{(d)} Corresponding PL and TRPL spectra at 290K (RT). At RT, PL spectra show increased noise, whereas the effect of temperature is not so prominently visible upon spontaneous emission decay. The unnormalized decay spectra in the inset show a reduction compared to the 10K data.}
    \label{fig:PL_TRPL}
\end{figure*}

Figure \ref{fig:FDTD_WL}.(a) presents differential reflectance data obtained from the bare HMM under broadband excitation taken at 4K and 300K.
Careful inspection of these spectra reveals the presence of closely spaced modes.
Single AgNW is capable of showing multiple Fabry-P\'erot resonances with its surface plasmons, closely spaced in frequency space, which was verified experimentally \cite{AgNW_FPresonance_PRL, AgNW_FPresonance}; in a recent theoretical report, a dielectric photonic crystal has indicated the presence of broad Fabry-P\'erot modes when it was in hyperbolic dispersion \cite{FPresonance_Hyerbolic_dielectric}. Therefore, it is imperative to look into mode behaviors theoretically, where the array effect for AgNWs and the presence of host Al$_2$O$_3$ could make this HMM a bit complex system regarding its multiple mode resonances.
\begin{table}[b]
\caption{\label{tab:HMM_WL_modes} Comparison for experimentally observed differential reflection data $\frac{\Delta R}{R}$ and numerically simulated data shown in Fig.\ref{fig:FDTD_WL}; analyses of emission center and corresponding linewidths are tabulated for modes around Rh6G emission at 4K and 300K.}
\begin{ruledtabular}
\begin{tabular}{ccccc}
\textrm{T}& \textrm{$\frac{\Delta R}{R}$}&\textrm{$\frac{\Delta R}{R}$}&\textrm{Simulated}&\textrm{Simulated}\\
&\textrm{Peak Centre}&
\textrm{Linewidth}&
\multicolumn{1}{c}{\textrm{Peak Centre}}&
\textrm{Linewidth}\\
\textrm{(K)}&
\textrm{(eV)}&
\textrm{(meV)}&
\multicolumn{1}{c}{\textrm{(eV)}}&
\textrm{(meV)}\\
\colrule
4 & 2.252 & 14.48 & 2.2541 & 10.44\\
& 2.271 & 13.76 & 2.2706 & 9.76\\
\colrule
300 & 2.2487 & 38.18 & 2.2489 & 36.44\\
& 2.2634 & 26.25 & 2.2696 & 28.73\\
\end{tabular}
\end{ruledtabular}
\end{table}
FDTD simulations of the model HMM system performed
with the commercially available package software \cite{Ansys} help in that purpose. The scattering field response of the periodic array HMM system was recorded, and corresponding field intensities $I_{Scat}$ are plotted in Fig.\ref{fig:FDTD_WL}.(a). The simulated data were overlaid upon the experimental data to better compare the mode properties obtained from differential reflection spectra. The data for material dielectric constants were taken from Johnson \& Christy's material data provided for the room temperature \cite{Johnson_Christy}. These functions determine the electromagnetic field responses in the simulation. Figure \ref{fig:FDTD_WL}.(b) shows the imaginary part of the dielectric constant of silver, which contains the effect of temperature; the real part of dielectric constants does not get modified. The scattering losses decrease as the temperature is reduced, and the effective damping parameter multiplies the available room temperature material data to yield material data at each relevant temperature, as presented here for 4K and the room temperature \cite{TempDepend_PL_Ag, AgNW_PhononScatt_SciRep}. For silver, $Re(\epsilon)\sim-12.86$ around the transition frequency of the emitter dye molecule presented in this work, which is about 2.25 eV, and that of alumina is taken at a constant value of 2.56. Around 4K, this damping constant is $\sim0.4$ times the room temperature value. The effect of the imaginary part of the material dielectric constants is seen in the scattering responses in Fig.\ref{fig:FDTD_WL}.(a).
\begin{table}[b]
\caption{\label{tab:Purcell_Expt} Summary of the observed Purcell enhancement and linewidth broadening at the lowest and highest temperature points in the experimental data shown in Fig.\ref{fig:PL_TRPL}.}
\begin{ruledtabular}
\begin{center}
\begin{tabular}{cccc}
\textrm{Temperature}&
\textrm{PL}&
\multicolumn{1}{c}{\textrm{Linewidth}}&
\textrm{Purcell}\\
\textrm{(K)}&
\textrm{Enhancement}&
\multicolumn{1}{c}{\textrm{Broadening}}&
\textrm{Factor}\\
\colrule
10 & 2.52 & 6.30 & 6.26\\
290 & 1.57 & 4.72 & 4.99\\
\end{tabular}
\end{center}
\end{ruledtabular}
\end{table}

Linewidths of the HMM modes can be estimated from the differential reflection and simulated scattering spectra, yielding the cavity mode decay parameter $2\kappa$. Numerically calculated spectra could resolve modes consistent with experiments; relevant spectral data for 4K and 300K peaks around Rh6G emission are discussed in Table \ref{tab:HMM_WL_modes}, where the numerical and experimental data are compared.

\begin{figure*}[!ht]
    \centering
    \includegraphics[width=1.0\linewidth]{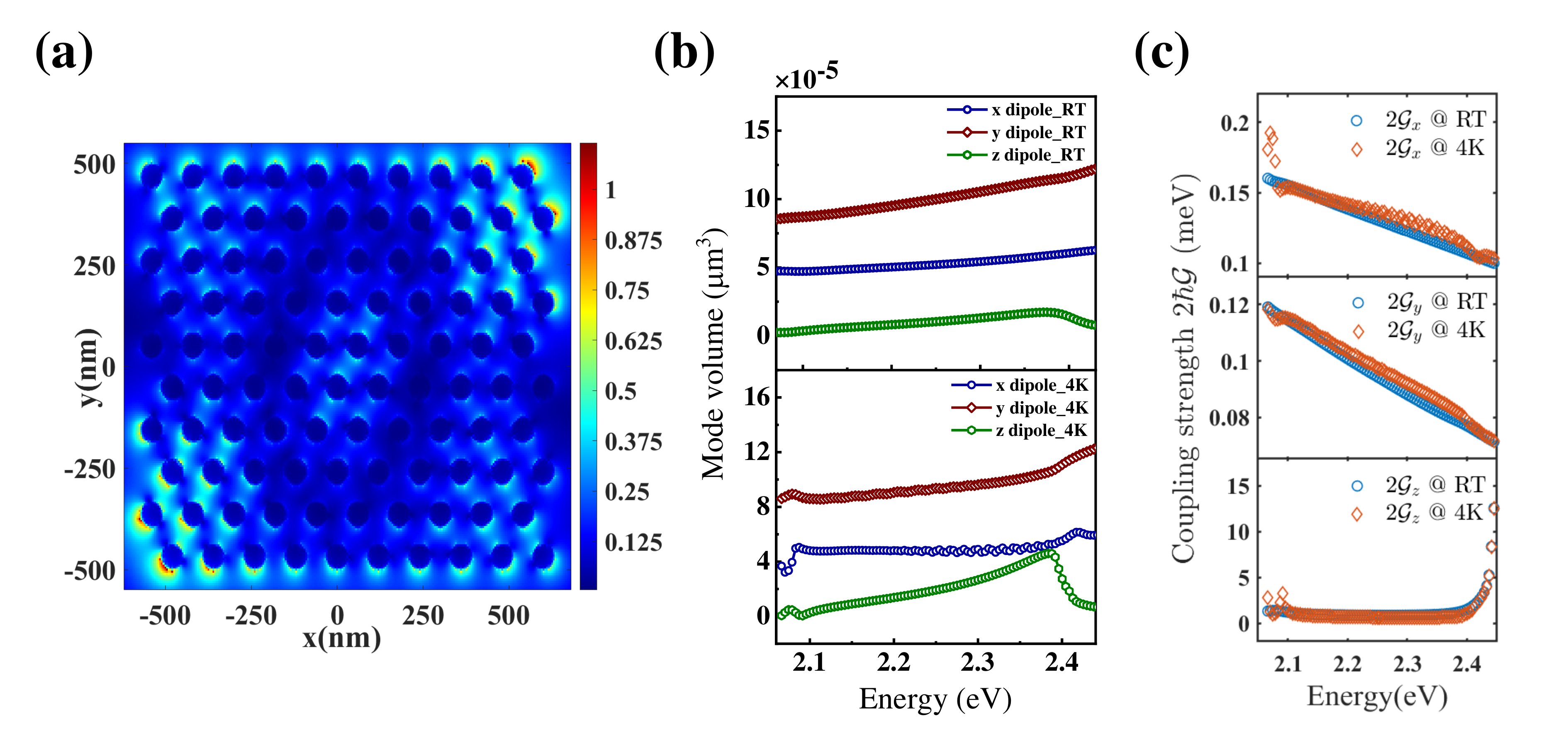}
    \caption{\textbf{(a)} Spatial field distribution of the local electric field at HMM interface. This plot is calculated at frequency $\sim$ 2.26 eV corresponding to the mode of scattering spectra as shown in Fig.\ref{fig:FDTD_WL}. 
    \textbf{(b)} Dispersion of the FDTD calculated HMM mode volume for dipole-excitation of different polarizations at RT (top panel) and 4K (bottom panel). \textbf{(c)} Dispersion of the coupling strength $2\hbar\mathcal{G}$ obtained using the mode volume dispersion data from (b).}
    \label{fig:modevol_coupling_g}
\end{figure*}

\subsection{Coupling of the cavity-like modes of HMM to quantum emitters}
We use Rhodamine 6G (Rh6G) dye as the quantum emitter. Rh6G is a high-quantum yield dye, shows an emission mode around $2.25$ eV in free space, i.e., safely located within the hyperbolic regime of this HMM, and doesn't have any dispersion of emission mode with temperature. Figure \ref{fig:PL_TRPL}.(a) \& \ref{fig:PL_TRPL}.(b) present the spectral data at 10K with prominent visibility due to reduced noise. Spontaneous emission decay spectra show an enhanced and faster decay, where the single exponent becomes a double exponent after dye emitters are coupled to the HMM modes. The effect of Purcell enhancement is seen in the PL spectra, where the linewidth broadening is more prominent than amplitude enhancement. Figure \ref{fig:PL_TRPL}.(c) shows increased noise in PL spectra at 290K (RT), and the effect of increased temperature is seen in the reduction of the maximum intensity of unnormalized spontaneous emission decay [Fig.\ref{fig:PL_TRPL}.(d)]; still, the decay rate is not modified much. To verify that easily, relevant data are compared in Table \ref{tab:Purcell_Expt}. 

Purcell enhancement is expected due to the modification of spontaneous emission decay rate $\gamma$ for the effect of modified PDOS, and it is expected to see an increment at the lowest temperature (10K) due to a decrease in scattering losses. Enhancement in the integrated PL intensity is complementary to support the Purcell enhancement. Since the decay rates of the HMM modes and the free-space spontaneous emission decay rate of Rh6G are much different in order of magnitude, the coupled system shows a bi-exponential decay due to the availability of two decay paths \cite{RKY_RTSinglePhoton_Plasmonic}. These two decay paths also affect the corresponding linewidths of the emission spectra of the coupled system. The slower path has a decay rate comparable to the free-space decay rate of the emitter, and the cavity-coupled path has a Purcell-enhanced faster decay rate. The corresponding linewidths of the emission spectra would reflect that; accordingly, the slower one has a linewidth comparable to that of free-space spectra, and the faster one has a broader linewidth. Due to this reason, PL spectra were fitted with a double-Lorentzian function to estimate these two different linewidths corresponding to the spectra of two available decay paths [See Appendix \ref{app:Double_Lorentz} for details]. After the fitting, broader linewidths of PL spectra were taken for further analyses. The broadening in Table \ref{tab:Purcell_Expt} only considers the broader linewidth of the fitted data.

\subsection{Estimate of the effective coupling constant}
Here, we discuss the theoretical analyses of the density matrices to arrive at the cavity properties of the HMM. The open quantum system of the dye emitter coupled to the HMM mode has three states that participate in the dynamics, $\ket{e,0}$,$\ket{g,1}$, and $\ket{g,0}$. Subsequently, the population dynamics come from Eq.(\ref{eq:master_eqn}) [See Appendix \ref{app:mastereqn} for details],

\begin{gather}\label{eq: main_density_matrix_1}
    \frac{\partial\rho_{11}}{\partial t} = -2\gamma_0\rho_{11}-i\mathcal{G}\rho_{12}+i\mathcal{G}\rho_{21},\\
    \frac{\partial\rho_{21}}{\partial t} = i\mathcal{G}\rho_{11}+i\delta\rho_{21}-(\kappa+\gamma_0+\Gamma)\rho_{21}-i\mathcal{G}\rho_{22},\label{eq: main_density_matrix_2}\\
    \frac{\partial\rho_{22}}{\partial t} = i\mathcal{G}\rho_{12}-i\mathcal{G}\rho_{21}-2\kappa\rho_{22}\label{eq: main_density_matrix_3}
\end{gather}
Here, $\rho_{ij}$ are the respective population density matrices for the states, $2\gamma_0$ is the spontaneous emission decay rate of the emitter in free space, $2\kappa$ is the photon decay rate from the cavity mode, $\Gamma$ is the dephasing rate of the emitter, and $2\mathcal{G}$ is the coupling strength for the emitter to the cavity mode.\\
To estimate the coupling parameter $2\mathcal{G}$, one needs the knowledge of $2\gamma_0$ and $2\kappa$. $2\gamma_0$ is extracted from the spontaneous emission decay measurements using TRPL, and $2\kappa$ is obtained from the differential reflection and simulated scattering spectra. We obtain the Purcell-enhanced modified spontaneous emission decay rate of the emitter coupled to the HMM from TRPL measurements. That decay rate is nothing but the constant of the exponent $e^{-t(a-\sqrt{b^2+c^2})}$, as obtained in Eq.(\ref{eq:rho11t}) and Eq.(\ref{eq:rho22t}) [See the details in Appendix \ref{app:mastereqn}].
A rigorous calculation to solve $2\mathcal{G}$ value starting from Eq.(\ref{eq: main_density_matrix_1}-\ref{eq: main_density_matrix_3}), with the assumption of adiabatic elimination of the faster variable applied to Eq.(\ref{eq: main_density_matrix_2}) \cite{adiabatic_elimin_1, adiabatic_elimin_2, RKY_RTSinglePhoton_Plasmonic}, since $\kappa$, $\Gamma \gg\gamma_0$, where the pure dephasing time for Rh6G dye is chosen to be $\sim$ 100 fs \cite{Rh6G_T2}, yields $2\mathcal{G} = 394.22\pm23.65$ GHz ($259.5\pm15.6\; \mu$eV), for the observed Purcell effect here [
c.f. Eq.(\ref{eq:rho11t})-(\ref{eq:rho_param_3})]. For simplicity, we made this estimate assuming the isotropic coupling of the dipole.

\subsection{Estimate of the mode volume from FDTD simulation}
With the help of FDTD simulations, we can independently verify the HMM system's cavity mode volume and coupling strength. FDTD simulation provides us with the distribution of the cavity field. Figure \ref{fig:modevol_coupling_g}.(a) presents the spatial distribution of the local electric field at the HMM interface; this field map is shown for the scattering mode $\sim 2.26$ eV at RT. 
Also, due to the anisotropy in HMM, FDTD can predict the cavity coupling strength and mode volume for each polarization.\\
The Quantum description of the electromagnetic field provides the electric field amplitude within a cavity \cite{GSA_QOpticsbook, S.Haroche_book},
\begin{equation}\label{eq:Q_Optics_Efield}
E = \sqrt{\frac{\hbar\omega}{2\epsilon\epsilon_0 V}}
\end{equation}
For cavity medium permittivity $\epsilon$ and mode volume $V$. A dipole emitter with transition electric dipole moment $|\Vec{\mu}|$ would experience the field with coupling strength $\hbar\Bar{\mathcal{G}}$,
\begin{equation}\label{eq:hg_strength}
    \hbar\Bar{\mathcal{G}} = |\Vec{\mu}|\sqrt{\frac{\hbar\omega}{2\epsilon\epsilon_0 V}}
\end{equation}
From Einstein A coefficients, we get \cite{EinsteinCoeff},
\begin{equation}\label{eq:EinsteinA}
    |\Vec{\mu}|^2 = \frac{3\epsilon_0hc^3}{2\omega^3}\gamma_0
\end{equation}
and,
\begin{equation}\label{eq:modevolume}
    V=\frac{\int \epsilon E^2d\mathcal{V}}{max(\epsilon E^2)}
\end{equation}
$\gamma_0$ is the vacuum spontaneous emission decay rate, $\omega$ is the angular transition frequency. Equation (\ref{eq:modevolume}) is used to define the mode volume in the FDTD simulation to calculate that numerically \cite{PNAS_HMM_indefiniteCavity, PRL_modevolume}; results are shown in Fig.\ref{fig:modevol_coupling_g}.(b). Such ultrasmall mode volume has recently been seen for other kinds of HMM systems \cite{SRK_ultrasmallHMM}, engaging in enhanced light-matter interactions. 
Hence, Eq.(\ref{eq:hg_strength}) gives the coupling strength, $2\hbar\mathcal{G}_x=135.04\;\mu$eV (4K) and $2\hbar\mathcal{G}_x=131.54\;\mu$eV (RT); $2\hbar\mathcal{G}_y=96.28\;\mu$eV (4K) and $2\hbar\mathcal{G}_y=94.48\;\mu$eV (RT); $2\hbar\mathcal{G}_z=683.48\;\mu$eV (4K) and $2\hbar\mathcal{G}_z=799.18\;\mu$eV (RT), for x, y and z polarizations, respectively of a single exciting dipole, and the dispersion is shown in Fig.\ref{fig:modevol_coupling_g}.(c). The coupling strength can be scaled with the number of dipoles as $\hbar\mathcal{G}\sim\sqrt{N}|\Vec{\mu}|E$ \cite{GSA_SC_N_emitters,P_Torma_SC_review}. An estimate can be made for $N$ here; 10$\mu$L Rh6G dye solution of 1.24 $\mu$M/mL concentration was added over the HMM sample. For this number concentration, obtained mode volume predicts $\sim$ 37 molecules available per mode, i.e., $\sim0.73$ molecules per (10 nm)$^3$ mode \cite{GSA_SC_HMM}, which falls short of achieving Rabi splitting. Therefore, results obtained from Eq.(\ref{eq: main_density_matrix_1}-\ref{eq: main_density_matrix_3}) and from Eq.(\ref{eq:hg_strength}) are in good agreement. 
Our complete set of results for cavity parameters of the HMM system, as obtained from solving master equations and FDTD simulations, is presented in Table \ref{tab:Result_compare}.
\begin{table}[!ht]
\caption{\label{tab:Result_compare} Summary of the cavity parameters obtained from master equation (ME) and simulation (FDTD) around the coupled emission transition frequency at room temperature. FDTD results are averaged over all dimensions. ME method takes the Q factor from the differential reflection data.}
\begin{ruledtabular}
\begin{center}
\begin{tabular}{cccc}
\textrm{Method}&
\textrm{Q factor}&
\multicolumn{1}{c}{\textrm{Mode volume}}&
\textrm{Coupling strength}\\
&
&
\multicolumn{1}{c}{\textrm{$\mu$m$^3$}}&
\textrm{($\mu$eV)}\\
\colrule
ME & 58.89 & 2.317$\times10^{-5}$ & 212.23\\
FDTD & 61.71 & 5.466$\times10^{-5}$ & 283.46\\
\end{tabular}
\end{center}
\end{ruledtabular}
\end{table}

\section{Discussion and Outlook}
In conclusion, this work has presented a comprehensive study of the light-matter interaction of the HMM system; first, from the differential reflection spectra of the HMM itself to identify its closely spaced modes, and then, it was coupled to a sparse number concentration of Rh6G dye to observe the Purcell effect in their weak coupling regime with the help of emission spectra and spontaneous emission decays. These results were used to quantify the coupling parameter from master equations. FDTD simulations helped complement and independently obtain the HMM mode information and extended its study to benefit from the anisotropic nature of metamaterials.
The proximity of the order of magnitudes between these two independent methods suffices the accuracy of our study regarding AgNW HMM. We can infer that ultrasmall mode volume is obtained and confines fields in subwavelength regimes, which is the key aspect of such plasmonic nanoscale metamaterials.
Therefore, AgNW HMM successfully shows cavity properties emerging as an indefinite medium cavity; these observed experimental results and theoretical analyses would further help to accurately estimate the polaritonic dynamics in the material. The ultrasmall mode volume within it provides the enduring effect of field confinement to become a viable photonic medium. Nonetheless, HMM is a lossy plasmonic material with damping effects that contribute significantly.  So, experiments at lowered temperatures were aimed to reduce the noise from scattering losses and extract the information more accurately. In that successful attempt, it is observed that HMM cavity properties do not get affected in the same order as individual AgNWs do due to damping, e.g., AgNW gets 2.5 times damping reduction \cite{AgNW_PhononScatt_SciRep} around 4K compared to RT; for the HMM, relevant results from experiments and FDTD simulations predict a variation of magnitudes $\sim 1.25 - 1.6$ and $\sim 1.27 - 2.6$, respectively. This poses an interesting gain for HMM as a plasmonic cavity medium. Significant observations with temperature variation are discussed in detail in our next work \cite{HMM2}. Moreover, based on our findings, we can predict the expected coupling strength for emitters of different oscillator strengths. For example, transition metal dichalcogenides (TMD) have very high oscillator strength; e.g., radiative linewidth $\sim 1.54$ meV \cite{KausikM_MoS2_linewidth}, and its transition dipole moment is mostly oriented in-plane \cite{MoS2_TDM_IPOP}. 
Therefore, Eq.(\ref{eq:hg_strength}) predicts $ 2\hbar\mathcal{G} \sim 18$ meV in case of in-plane coupling to HMM, 
and $2\hbar\mathcal{G}\sim120$ meV, if out-of-plane coupling was possible with HMM for single dipoles at RT.

\section{Acknowledgement}
J.K.B. acknowledges funding from the Anusandhan National Research Foundation (ANRF), India, through grant number CRG/2021/003026 and DST, FIST grant. G.S.A. thanks the support of the INFOSYS CHAIR Visiting Professorship of IISc, which made this collaboration possible. G.S.A. also thanks the Robert A. Welch Foundation (Grants No. A-1943-20240404)
for supporting this work. A.N. thanks the Advanced Facility for Microscopy and Microanalysis (AFMM),
Indian Institute of Science, Bangalore, for access to SEM
and FIB facilities. A.N. thanks Harshavardhan R.K., a former graduate student in the Dept. of Physics, Indian Institute of Science, for useful discussions and suggestions for setting up the experiment.

\onecolumngrid
\appendix
\section{Derivation of coupling strength from Master equations}\label{app:mastereqn}
We start with describing the Hamiltonian of the open quantum system of dye-coupled to HMM.
Coupling parameter $\mathcal{G}$ of the light-matter interaction is defined in the Hamiltonian \cite{GSA_mastereqn_book, 
RKY_RTSinglePhoton_Plasmonic},
\begin{equation}\label{eq:Coupled_Hamiltonian}
    H = \hbar\omega_0 \hat{S}_z + \hbar\omega_m \hat{b}^\dag \hat{b} + \hbar \mathcal{G}(\hat{S}^+\hat{b} + \hat{S}^-\hat{b}^\dag)
\end{equation}
where $\omega_0$ is the dipole transition frequency of the dye emitter, $\omega_m$ is the resonant frequency of the HMM mode, and $\hat{b}^\dag (\hat{b})$ defines the raising (lowering) operators for the HMM mode $m$.
Here, the dye emitter is considered as a simple two-level system with ground state $\ket{g}$, and excited state $\ket{e}$. Spin-$\frac{1}{2}$ operators are defined for the two-level system dipole transitions; 
\begin{gather}\label{eq:S_operators}
    \hat{S}^+ = \ket{e}\bra{g},\\
    \hat{S}^- = \ket{g}\bra{e},\\
    \hat{S}_z = \frac{1}{2}(\ket{e}\bra{e} - \ket{g}\bra{g})
\end{gather}

Dynamics of the emitter, HMM modes, and the coupled system can be estimated from the master equation \cite{RKY_RTSinglePhoton_Plasmonic},
    \begin{gather}\label{eq:master_eqn}
        \frac{\partial\rho}{\partial t} = -\frac{i}{\hbar}[H',\rho] - \kappa(\hat{b}^\dag \hat{b}\rho - 2\hat{b}\rho\hat{b}^\dag + \rho\hat{b}^\dag \hat{b})-\gamma_0(\hat{S}^+ \hat{S}^-\rho -2\hat{S}^-\rho\hat{S}^+ + \rho\hat{S}^+ \hat{S}^- )-\Gamma(\hat{S}_z\hat{S}_z\rho -2\hat{S}_z\rho\hat{S}_z + \rho\hat{S}_z\hat{S}_z),\\
        H' = \hbar\delta\hat{b}^\dag \hat{b} + \hbar\mathcal{G}(\hat{S}^+\hat{b} + \hat{S}^-\hat{b}^\dag), \;\,\, \delta = (\omega_m - \omega_0) \label{eq:RWA_Hamiltonian}
    \end{gather}

defined in the rotating frame of the emitter transition frequency $\omega_0$, and $\Gamma$ is the dephasing rate of the emitter. Three states of the coupled system are participating in the dynamics, $\ket{e,0}$,$\ket{g,1}$ and $\ket{g,0}$. Subsequent population dynamics comes from Eq.(\ref{eq:master_eqn}),

\begin{gather}\label{eq:density_matrix_1}
    \frac{\partial\rho_{11}}{\partial t} = -2\gamma_0\rho_{11}-i\mathcal{G}\rho_{12}+i\mathcal{G}\rho_{21},\\
    \frac{\partial\rho_{21}}{\partial t} = i\mathcal{G}\rho_{11}+i\delta\rho_{21}-(\kappa+\gamma_0+\Gamma)\rho_{21}-i\mathcal{G}\rho_{22},\label{eq:density_matrix_2}\\
    \frac{\partial\rho_{22}}{\partial t} = i\mathcal{G}\rho_{12}-i\mathcal{G}\rho_{21}-2\kappa\rho_{22}\label{eq:density_matrix_3}
\end{gather}

Eq.(\ref{eq:density_matrix_2}) is equaled to zero under the adiabatic elimination of faster variables \cite{RKY_RTSinglePhoton_Plasmonic,adiabatic_elimin_1,adiabatic_elimin_2}, which yields
\begin{equation}\label{eq:mastereq_adiab}
    \rho_{21}=\frac{i\mathcal{G}}{\kappa+\gamma_0+\Gamma}(\rho_{11}-\rho_{22})
\end{equation}
Putting Eq.(\ref{eq:mastereq_adiab}) in Eq.(\ref{eq:density_matrix_1}) and Eq.(\ref{eq:density_matrix_3}) gives
\begin{gather}
    \dot{\rho}_{11}=-2\gamma_0\rho_{11}-\frac{2\mathcal{G}^2}{\kappa+\gamma_0+\Gamma}(\rho_{11}-\rho_{22}),\\
    \dot{\rho}_{22}=-2\kappa\rho_{22}+\frac{2\mathcal{G}^2}{\kappa+\gamma_0+\Gamma}(\rho_{11}-\rho_{22})
\end{gather}
Taking the Laplace Transform of these two equations give,
    \begin{gather}\label{eq:LT_1}
    sP_{11}(s)-\rho_{11}(0)=-2\left(\gamma_0+\frac{\mathcal{G}^2}{\kappa+\gamma_0+\Gamma}\right)P_{11}(s)+\frac{2\mathcal{G}^2}{\kappa+\gamma_0+\Gamma}P_{22}(s),\\
    sP_{22}(s)-\rho_{22}(0)=-2\left(\kappa+\frac{\mathcal{G}^2}{\kappa+\gamma_0+\Gamma}\right)P_{22}(s)+\frac{2\mathcal{G}^2}{\kappa+\gamma_0+\Gamma}P_{11}(s)\label{eq:LT_2}
\end{gather}
with the initial conditions are $\rho_{11}=1$ and $\rho_{22}=0$. These lead Eq.(\ref{eq:LT_1} \& \ref{eq:LT_2}) to
    \begin{gather}\label{eq:P22_LT}
    P_{22}(s)=\frac{\frac{2\mathcal{G}^2}{\kappa+\gamma_0+\Gamma}}{s+2\left(\kappa+\frac{\mathcal{G}^2}{\kappa+\gamma_0+\Gamma}\right)}P_{11}(s),\\
    P_{11}(s)\left[s+2\left(\gamma_0+\frac{\mathcal{G}^2}{\kappa+\gamma_0+\Gamma}\right)-\frac{\frac{4\mathcal{G}^2}{\kappa+\gamma_0+\Gamma}}{s+2\left(\kappa+\frac{\mathcal{G}^2}{\kappa+\gamma_0+\Gamma}\right)}\right]=1\label{eq:P11_LT}
\end{gather}
When $\mathcal{G}=0$, $P_{11}(s)=\frac{1}{s+2\gamma_0}$ and $P_{22}(s)=0$, i.e., the results retrieve the uncoupled dynamics.\\ 
After rigorous calculations to simplify Eq.(\ref{eq:P11_LT} \& \ref{eq:P22_LT}),  these are arranged in the form of known Laplace Transforms,
\begin{gather}
    \frac{s+a}{(s+a)^2-b^2}\rightarrow e^{-at}\cosh{bt},\\
    \frac{b}{(s+a)^2-b^2}\rightarrow e^{-at}\sinh{bt}
\end{gather}
Accordingly, applying inverse Laplace transform on Eq.(\ref{eq:P11_LT} \& \ref{eq:P22_LT}) gives general solutions of state populations,
    \begin{gather}\label{eq:rho11t}
        \rho_{11}(t)=\frac{1}{2}\left[\left(1+\frac{b}{\sqrt{b^2+c^2}}\right)e^{-t(a-\sqrt{b^2+c^2})}+\left(1-\frac{b}{\sqrt{b^2+c^2}}\right)e^{-t(a+\sqrt{b^2+c^2})}\right],\\
        \rho_{22}(t)=\frac{c}{2\sqrt{b^2+c^2}}\left[e^{-t(a-\sqrt{b^2+c^2})}-e^{-t(a+\sqrt{b^2+c^2})}\right]\label{eq:rho22t}
    \end{gather}
Where, 
\begin{gather}\label{eq:rho_param_1}
    \frac{2\mathcal{G}^2}{\kappa+\gamma_0+\Gamma}+\kappa+\gamma_0=a,\\
    \kappa-\gamma_0=b,\\
    \frac{2\mathcal{G}^2}{\kappa+\gamma_0+\Gamma}=c\label{eq:rho_param_3}
\end{gather}
were chosen for ease of expressing the equations \ref{eq:rho11t} \& \ref{eq:rho22t}. Here, the value of $2\gamma_0\sim0.34\times10^9$ s$^{-1}$, and $2\kappa$ corresponds to the relevant HMM mode linewidths in Table \ref{tab:HMM_WL_modes}. From the TRPL data in Fig.\ref{fig:PL_TRPL}, the maximum decay rate of faster decay path is found to be $\sim3.5\times10^9$ s$^{-1}$. Among the two decay rates in Eq.\ref{eq:rho11t} \& Eq.\ref{eq:rho22t}, the decay constant of the exponent $e^{-t(a-\sqrt{b^2+c^2})}$ lies in the given frequency range of the observed spontaneous emission decay rates, which can be verified for the decay rates given. For example, at 4K, for the value of $2\kappa$ obtained from Table \ref{tab:HMM_WL_modes}, $a+\sqrt{b^2+c^2}\sim4\kappa$, and $\frac{b}{\sqrt{b^2+c^2}}\sim0.99$, therefore, the coefficients of $\rho_{11}(t)$ are 0.995 and 0.005, respectively, and that for $\rho_{22}(t)$ is $1.447\times10^{-4}$. This example justifies the observations in the experiment. Thus, we obtain the coupling strength parameter 2$\mathcal{G}=394.22\pm23.65$ GHz, as shown in the main text, for the given values of $2\gamma_0$ and $2\kappa$.
\section{Double Lorentzian Fitting of PL spectra}\label{app:Double_Lorentz}
We have used the double Lorentzian function to fit the PL spectra of the dye emitter emission coupled to the HMM system. This function is described as
\begin{equation}
    f = A_1\frac{\frac{\Gamma_1}{2}}{(E-E_0)^2+(\frac{\Gamma_1}{2})^2}+A_2\frac{\frac{\Gamma_2}{2}}{(E-E_0)^2+(\frac{\Gamma_2}{2})^2}
\end{equation}
where $E_0$ is the emission peak's central frequency for both spontaneous emission decay paths, $\Gamma_1$ and $\Gamma_2$ are corresponding decay linewidths. Since the decay rates of the HMM modes and the free-space spontaneous emission decay rate of Rh6G are much different in order of magnitude, the coupled system is expected to show bi-exponential decay due to the availability of two decay paths \cite{RKY_RTSinglePhoton_Plasmonic}. These two decay paths also affect the corresponding linewidths of the emission spectra of the coupled system. The slower path has a decay rate comparable to the free-space decay rate of the emitter, and the cavity-coupled path has a Purcell-enhanced faster decay rate.  Accordingly, the linewidths of the emission spectra would reflect the presence of these two decay paths, the slower one having a linewidth comparable to that of free-space spectra and the faster one having a broader linewidth. Due to this reason, PL spectra were fitted with a double-Lorentzian function to estimate these two different linewidths corresponding to the spectra of two available decay paths. After the fitting, the broader linewidth of the PL spectra was taken for further analyses, and the linewidth broadening presented in Table \ref{tab:Purcell_Expt} only considers the broader linewidth of the fitted data.
\section{HMM fabrication and the Measurement set up}\label{app:setup}
AgNW HMM is fabricated in the porous alumina matrix by electrochemically depositing the AgNWs through the pores of the matrix, following the well-established fabrication procedure previously followed in the group \cite{HMM_Harsha, HMM_RKY, HMM_RKY2, HMM_SRK}. The thin film of porous alumina matrix is prepared by two-step anodization of highly pure, commercially available Aluminum metal sheet procured from Alfa Aesar. The metal piece is electropolished and anodized in a dilute acidic medium. The first anodization step takes 14 hours, and the set voltage is 40V. The set voltage for the anodization determines the porosity of the matrix, which is also the same as the metal filling fraction in the HMM, $f$; here, $f$ is $\sim20\%$. It is given in terms of the nanowire diameter $d$ and their periodicity $a$ as,
\begin{equation}
    f=\frac{\pi}{2\sqrt{3}}\left(\frac{d}{a}\right)^2
\end{equation}
The anodization timing of the second step determines the thickness of the alumina matrix grown over the metal. After the anodization, the bottom metal and the barrier layer of alumina are removed with chemical etching procedures. Once the thin film of alumina matrix is received, it is sputtered with a thin gold film acting as a nano-electrode and used to electrodeposit AgNWs through the pores in a silver bromide (AgBr) solution. After electrodeposition, the top metal layer of silver is etched out with the Focused Ion Beam (FIB) etching, and using the SEM, the HMM sample is imaged (Fig.\ref{fig:schematic}).\\   
Once the HMM is fabricated and the sample for the measurement is prepared, dispersing the Rh6G dye emitters over it, PL and TRPL measurements are performed with a confocal microscope (WiTec Alpha 300) with 50$\times$ magnification and 0.55 N.A. objective, connected to the CCD for PL signal collections and APD for TCSPC measurements. HMM sample was attached over a Si-SiO$_2$ substrate; the same substrate was used for control measurements with the dye. Samples were placed inside the vacuum chamber of the MicrostatHe cryostat (Oxford instruments). A white-light lamp source and a pulsed-to-CW convertible 405 nm emission diode laser (PicoQuant PDL 800-D) were used for broadband and PL, TRPL measurements, respectively.
Broadband differential reflection measurements were performed on the HMM sample with the help of the white-light lamp source, and the same measurement setup was kept to collect the scattering responses from the HMM in an index-matched ambience before the dye was deposited.

\FloatBarrier
\twocolumngrid
\bibliography{HMMcavity}

\begin{thebibliography}{60}
\bibitem [{HMM()}]{HMM2}%
  \BibitemOpen
  \href@noop {} {}\bibinfo {note} {See the Article Letter at URL-will-be-inserted-by-publisher for the detailed results and discussion on the effect of Temperature on HMM as an Indefinite Plasmonic Cavity Medium.}\BibitemShut {Stop}%
\end{thebibliography}

\end{document}